# IoT-based Efficient Streetlight Controlling, Monitoring and Real-time Error Detection System in Major Bangladeshi Cities

A.T.M Mustafa Masud Chowdhury
*Department of Computer Science and Information Technology*
*Southern UniversityBangladesh*
Chattogram, Bangladesh

Jeenat Sultana
*Department of Computer Science and Information Technology*
*Southern UniversityBangladesh*
Chattogram, Bangladesh

Md Sakib Ullah Sourav
*School of Management Science and Engineering*
*Shandong University of Finance and Economics*
Jinan, China

*Abstract*—Meeting the increasing electricity demand has become a major issue for every nation, especially for a third world country like Bangladesh. A huge wastage of electricity can be seen in Bangladesh due to improper street light management that leads to an enormous financial loss every year. Many noteworthy works have been done by researchers from different parts of the world in tackling this issue by using the Internet of Things (IoT) yet very few in Bangladesh's perspective. In this work, we propose an efficient IoT-based integrated streetlight framework that offers cloud powered monitoring, controlling through light dimming as per external lighting conditions and traffic detection, as well as fault detecting system to ensure low power and electricity consumption. We analyzed data from Dhaka North and South City Corporation, Narayanganj City Corporation and Chattogram City Corporation where our proposed model demonstrates a reduction in energy cost of up to approx. 60% than that of the existing system. This intelligent and efficient system can be implemented in major cities of Bangladesh as part of a smart city and the surplus of the electricity can be utilized for household use and other essential purposes.

## I. INTRODUCTION

The discovery of fire was crucial to the advancement of human civilization through which men is set free from the clutches of darkness. This has an influence on the human race as a whole and improves its performance. This is something we see in cities. The major distinction between town and village is the presence of street lights, which brighten city streets by removing darkness. The world's first street light was installed in London in the winter of the year of 1417. Later got introduced in the United States for the first time by Benjamin Franklin, who reproduced his own version of the glass globes used in London, finally the Pennsylvania Assembly Circa 1757 oil candles [1]. Street light system was considerably enhanced with the discovery of electricity. Yablokov candles are a type of electric carbon arc lamp created by Pavel Yablokov, a Russian electrical engineer, in 1876. The world's first electric streetlight is said to be in Paris, France on May 30, 1878 [2]. As a result, we may deduce that street lights are an important part of a city's infrastructure.

The internet and information technologies have ushered in a new era. The Internet of Things (IoT) connects each device to the Internet (devices that can connect to the Internet), and one device completes an integrated job with another device via data transmission [3]. Home automation is one example of successful implementation of IoT. Furthermore, automation is a significant part of modern society and our everyday lives. As a result, we can observe that the old systems that surround us are rapidly altering to make our lives simpler. We have seen how much energy, particularly electricity, is wasted in Bangladesh's major cities. According to a report, streetlights account for 30%-35% percent of a city's overall power usage [4]. Even in this age of information technology, streetlights in Bangladesh are turned ON and OFF in a conventional manner. As a result, from the investigation on Chattogram streets we observe that the streetlights are kept turned on till noon 12 p.m. We've concluded that relying on technology to make Bangladesh's cities green and sustainable would be necessary, thus we're aiming to build green and smart cities through IoT.

To save money on electricity, we are proposing an IoT-based streetlight monitoring, control system as well as cost monitoring, immediate defective light detection, and automatic dimming of streetlights with 50% brightness when traffic congestion is decreased. The ultra-sensor recognizes when a vehicle or a person passes a street light electric pole, and the lights will shine at full brightness at that time.

The rest of the paper is structured as follows. Section 2 discusses the best practices of modern streetlight models and systems as literature review. Section 3 highlights the methodology of our proposed model. In section 4, we focus on the features that reveal the validity of the proposed model of this work. We depict the results and discussion in Section 5. Finally, we conclude the paper in section 6.

## II. LITERATURE REVIEW

Hannan et al. in their both works [5, 6] developed a LDR and ultrasonic sensors-based street lighting system based on real-time data where they did not mention in which method they collected data while analyzing their results and where to store real-time data in their prototype. Alex et al. [7] used ZIGBEE and sensors to investigate an energy efficient intelligent street lighting system. To improve the efficiency of the street lighting system, ZIGBEE and sensors were used. ZIGBEE and sensors are responsible for the system's low power usage. The communication range of ZIGBEE, on the other hand, is around 50 meters. As a result, it is inconvenient to use.

Bhairi et al. [8] looked into smart solar Light Emitting Diode (LED) streetlight which is programmed to turn off automatically during the day and only turn on at night. During severe rains or terrible weather, the light will glow at 30% brightness, but if there is a person or automobile nearby, it will illuminate at 100% brightness. Here, there isn't any form of error-detection mechanism.

Few studies [9-12] analyzed the LoD (Light of Demand) mechanism with motion detectors, intensity control systems, AC voltage regulators and controllers, including AC-AC buck converters, multi-taped autotransformers, and high- (Fig.1.a, and b). A detailed documentation of our investigation covering various national news articles, media reports and our field survey has been added to Appendix I for better understanding of the readers of this study.

Light Dependent Resistor (LDR) on our proposed system which will burn automatically when there is no sunlight and frequency switch converter dimmer systems. The solutions presented in these studies for decreasing LED streetlights. Although these methods were successful in dimming LED lighting, they all necessitated the purchase of expensive materials.

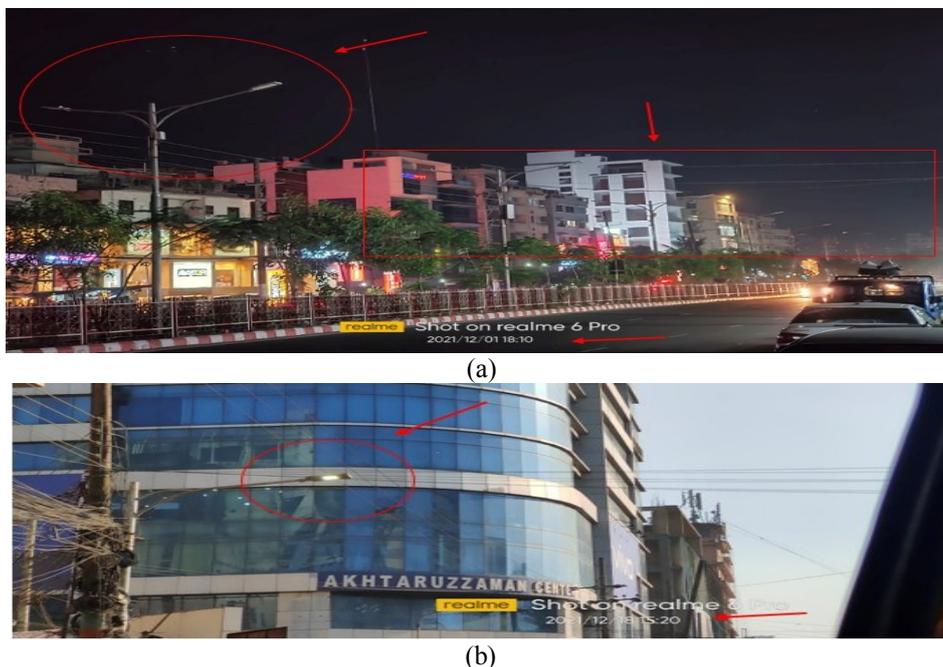

**Fig.1.** The mismanagement of streetlight controlling at Chittagong from our investigation, (a) night-time, (b) day-time.

IoT-based street lighting and traffic control systems were investigated by Saifuzzaman et al [13]. The notion of merging road traffic systems with street lighting is discussed. However, it is less relevant in Bangladesh due to a lack of coordination among Bangladeshi governmental entities. The Metropolitan Police Department is in charge of traffic, while the city corporation is responsible for street lighting. Additionally, the model proposed in [14] has introduced computer vision-based streetlight controlling system which is quite new in Bangladesh's perspective and didn't provide a holistic framework of streetlight monitoring, controlling and error detecting at the same time that we focused in this work.

Each of the studies listed above was outstanding. Researchers attempted to provide the most efficient methods of their time. However, none of the works amalgamate all the four features of monitoring, fault detecting, brightness controlling and low power consumption. Thus, our focus is to study and assess the potential of IoT in prevention the wastage of money through efficient streetlight controlling system in Bangladesh. Chattogram, Bangladesh's second-largest city, where we live, we studied and uncovered the reality, a glimpse of that can be shown in Fig.1. We obtained this data by observing in Port Connecting Road and Agrabad Access Road in Chittagong city from 5:25 p.m. to 8:30 p.m. on November 28th and December 1st, 2021 will turn off automatically when it gets sunlight. In the conventional method, the authorities have no way of knowing whether any of the city's street lights are faulty. However, our model is capable of detecting real-time problems. Controlling the brightness of street lights can also help to save power.

### III. METHODOLOGY

Our system is developed on two primary components: hardware and software (admin panel). Here, we'll take some ideas about our system's hardware and use a block diagram and flow chart to try to understand how our system works by integrating the software with the hardware.

#### A. Hardware Specifications

The Arduino Nano served as the system's hardware brain. The ATmega 326 microcontroller has 14 digital pins, 8 analog pins, and a clock speed of 16 MHz. We used ultrasonic sensors to detect the presence of any vehicle or person on the road. The ultrasonic sensor has 4 pins, positive, negative, trigger and echo. LDR is used to see the brightness of the street light and identify the presence of sunshine. We've utilized esp8266 as a Wi-Fi module to link our hardware system to the control room over the internet. LED bulbs are the most modern bulbs at present. It saves 75% of the energy used by conventional lights and lasts 25 times longer. We used current sensors to monitor AC and DC currents and offer electrical isolation between the sensor's output and the circuit being examined & potential meter that divide a higher voltage by a predetermined ratio based on the electrical components to produce a lower output voltage. We used some more components those are crucial to our system, but the above are the most significant.

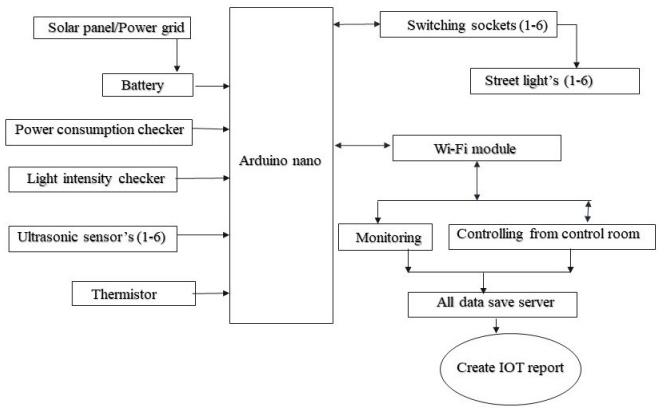

Fig.2. Block diagram of the system we proposed.

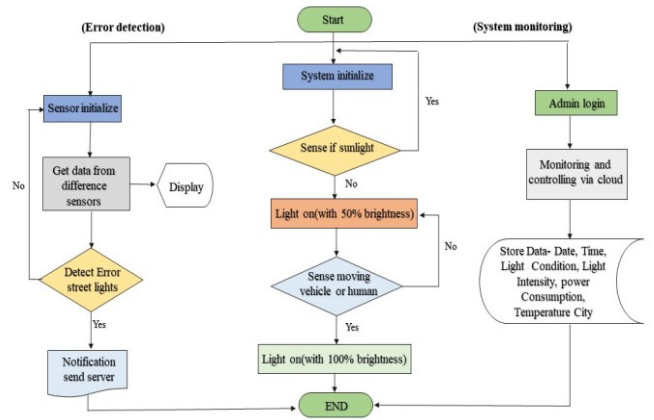

Fig.3. Flow chart of the system we proposed.

B. *Block Diagram and Flow Chart of the model*

As seen in the block diagram, the Arduino Nano is the central system or brain. By charging the battery through a solar or grid, it provides a power supply input as well as a power consumption checker, light intensity checker, ultrasonic sensor (No.1-6), and thermistor input data to Arduino Nano. The Arduino Nano, on the other hand, collects all of that data and controls the bulbs using switching sockets.

Additionally, the data is sent to the system for monitoring through Wi-Fi. If necessary, it collects input from the control room and delivers it to the Arduino Nano through Wi-Fi. The Arduino Nano re-gives the streetlights instructions from the control room via a switching connector. An IoT report is made by transmitting all of the data servers and employing the most up-to-date monitoring and regulating. We've explained how our hardware works using the block diagram in Fig 2. Fig.3 explains on how hardware and software interact and function together as flowchart.

Our system works in three ways in parallel with the one we started. The right-hand side of the screen is devoted to data collecting and system monitoring. Error Detection is located on the left side of the screen. We will start with system monitoring. We have a system admin panel that will display all of our system's information when signing in and gathering the required data. An emergency control system is accessible from here. In the second section, we'll talk about our system error detection work.

The program will monitor a variety of hardware sensors in the system we've developed. If the system identifies a streetlight malfunction, our server will send a notification, and if no problem is identified, the monitoring will continue as long as the straight light is on. in the third level, our system will measure the quantity of sunlight before turning on the street light. If the quantity of sunlight is minimal or nonexistent, the street light will be turned on; if there is sunlight, it will not be turned on. the next step is to see if there are any cars or people around our system street lights. If there are people or cars, our system will direct the street lights to turn on with 100% brightness if there are no people or cars around the street lights and the street lights are at night. Instruct to start with 50% brightness. It will, however, do its duty. After our given period, the 50% brightness scheme will take effect. This method will be used to complete all three forms of work or stages.

## IV. FEATURES OF THE PROPOSED MODEL

Our lab prototype was designed to tackle a wide range of real-world issues. Features that our proposed model possess are-

*Sunlight diagnosis.* In the experiment, we first tested whether our system can detect sunlight.

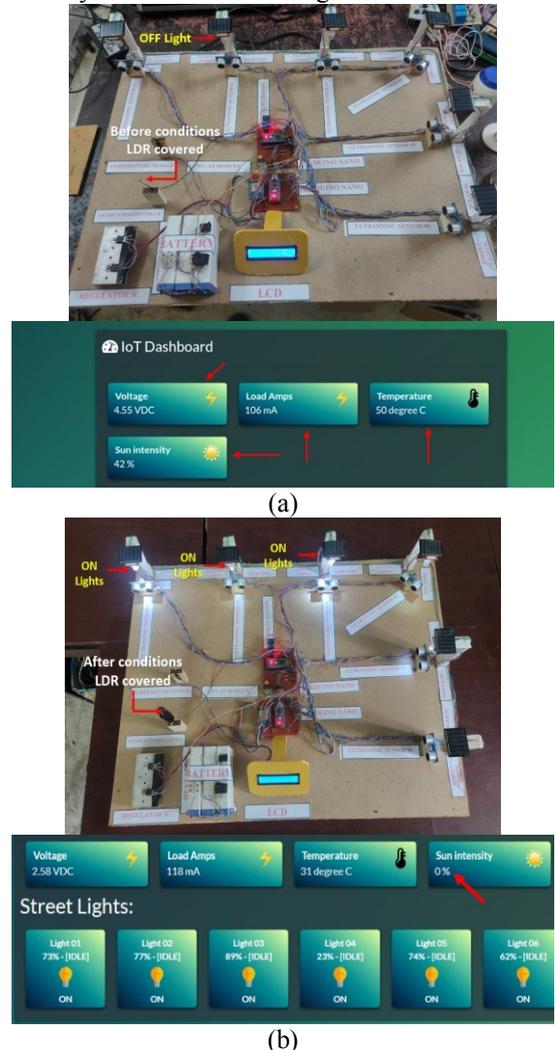

Fig.4. Sunlight detection with the help of LDR

In Fig.4 (a), we can see the LDR sensor can detect the lab's electric light, as seen in the model in Fig.4 (a). As a result, all of our model's streetlights have been turned off. Also we can see that our admin panel is displaying the intensity of the sunlight, which is at 42% to create an artificial night. Subsequently, we covered the LDR with a pen cap in the Fig

4 (b). In the admin panel, we can see that the sun's intensity is now set at 0 percent and all of our street lights have been successfully turned on automatically. We can also see the brightness of each light in the admin panel.

*Determine real-time current consumption of all sensors.* Fig.5 shows how much electricity is consumed when our system is turned on, as well as the current weather temperature.

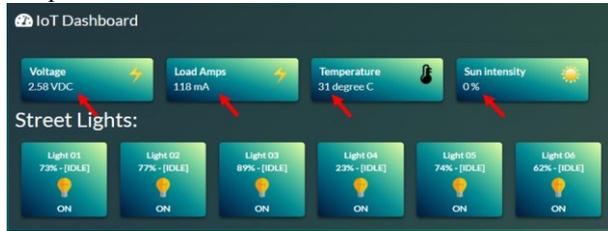

**Fig.5.** Electricity consumption data in admin panel

*Real-time error detection.* Because we conducted this experiment in the lab, our system artificially manufactured an error. Looking at our model and admin panel, we can see the condition before artificially producing the error in the Fig.6 (a). We are covered the LDR sensor on our Street Light 3 to produce a fake error when the LDR sensor can't detect the street light. Then our admin panel will see that the system has successfully demonstrated that street light number 3 is damaged, yet the light is on in Fig.6 (b).

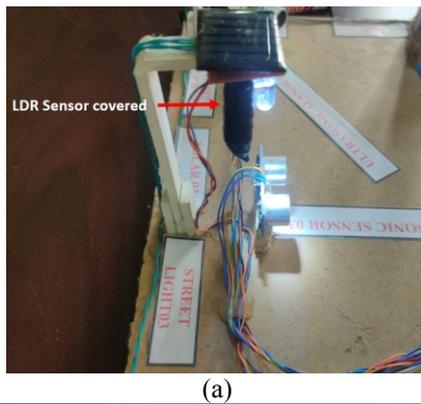

(a)

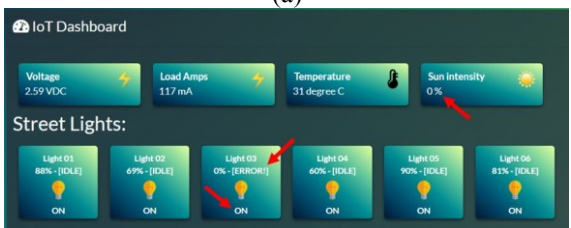

(b)

**Fig.6.** Light 3 has turned on after LDR sensor is artificially covered (a) and shown in admin panel (b)

*Testing ultrasonic sensors.* Ultrasonic sensors are used in our system to detect if there aren't any cars in the region of the road lights. If a car approaches, the brightness of the street lights in that zone will significantly rise. This can be seen in the Fig.7. When there were no automobiles, the street lights were turned off. As we can observe in the admin panel, when we park a bus in front of ultrasonic, the brightness of the street lights is substantially higher than other street lights (Fig.7.b). In this way, it is possible to reduce power waste.

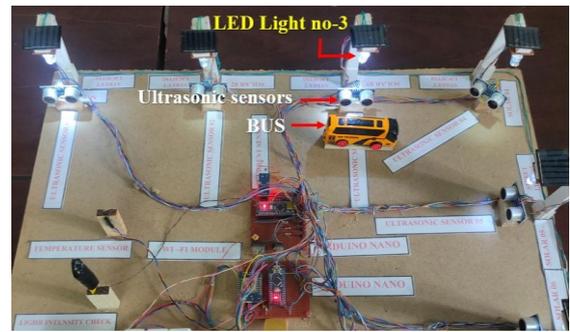

(a)

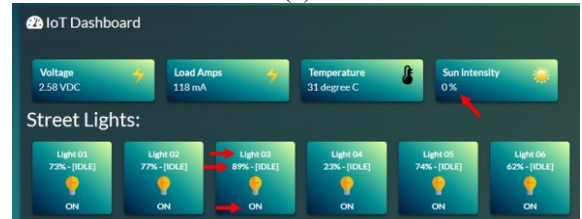

(b)

**Fig.7.** Ultrasonic sensor in Light 3 detects a car (a) and the brightness of the Light 3 has instantly increased (b)

*Data storing.* Our system can record all of the information in the form of tables, including date and time, as shown in the Fig.8 so that we may do additional studies on this in the future.

**Fig.8**. Data of streetlights in admin panel

## V. RESULT & DISCUSSION

Following our lab experiment, the focus is on attempting to shed light on how to save money and power by replacing the traditional street lighting system with the system we have constructed. From the Chittagong City Corporation's (CCC) website, description of different types of streetlights, namely, sodium lights, 100 watts bulbs, halogen bulbs, metal halide bulbs, tube lights, totaling 33,750 incandescent street lights can be seen [15]. As well as, Dhaka North City Corporation (DNCC), Dhaka South City Corporation (DSCC), Narayanganj City Corporation (NCC) has 46,410, 54,966 and 2,474 LED streetlights respectively [16, 17, 18]. We calculate and compare energy consumption and energy cost in these four city corporations in three scenarios.

**Scenario 1:** We estimated all of CCC's 33,750 streetlights at 100 watts (W) sodium lights to see how much the existing system costs. Most of the time, streetlights are used for 12 hours a day.

For a 100W sodium streetlight, consumption for 12 hours is 1.2 kWh. Per unit cost is 7.70 TK/kWh [12]. So daily Electricity cost for a single 100W sodium streetlight is (1.2 kWh*7.70) = 9.24 TK

Hence, total energy consumption is = (1.2*33750) = 40500KW = **40.5MW**.

None of DNCC, DSCC and NCC has sodium lights as streetlight in their existing system. Hence, Scenario 1 is not applicable for these three city corporations.

**Scenario 2:** if we replace the existing 100W sodium lights of CCC with 40W LED lights, consumption of power for 12 hours will be 0.48 kWh and daily electricity cost is 3.70 TK. Hence, total energy consumption with 100% brightness,
For 33750 no's LED light in CCC is **16.2 MW**
For 46410 no's LED light in DNCC is **22.3 MW**
For 54966 no's LED light in DSCC is **26.4 MW** and
For 2474 no's LED light in NCC is **1.2 MW**

**Scenario 3:** As our model would dim the brightness of LED streetlights in absence or less congestion of traffic, say for example, lights would be in dimming condition on 50% of its overall active time as traffic congestion is lesser during the midnight to dawn hours. For a 40W LED Streetlight in CCC, considering 50% of dimming condition, the power consumption of a light should be half, which is approx. 20W per light. Hence, consumption for 12 hours is 0.24 kWh, daily Electricity cost is 1.85 and the total energy consumption with 50% brightness,
For 33750 no's LED light in CCC light is **8.1 MW.**
For 46410 no's LED light in DNCC is **11.15 MW**
For 54966 no's LED light in DSCC is **13.2 MW** and
For 2474 no's LED light in NCC is **0.6 MW**

Table 1: Energy consumption comparison in CCC

| Lighting option | Per day (TK) | Per month (TK) | Per year (TK) |
|---|---|---|---|
| For 100W sodium light | 311850 | 9355500 | 113825250 |
| For 40W LED light (with 100% brightness) | 124875 | 37 46250 | 45579375 |
| For 40W LED light (with 50% brightness) | 62438 | 1873125 | 22789688 |
| Total Save [100W - 40W] (100% brightness) | 186975 | 5609250 | 68246125 |

From Table 1, we can see the comparison between electricity consumption among the existing 100W sodium street lighting system and two other discussed options in CCC in accordance with our proposed model.

Table 2: Energy cost comparison in CCC

| For 100W incandescent light/day | For 40W LED light (with 100% brightness)/day | For 40W LED light (with 50% brightness)/day |
|---|---|---|
| 40.5 MW | 16.2 MW | 8.1 MW |

In Table 2, we showed the corresponding energy costs for all the three scenarios discussed-above in terms of day, month and year.

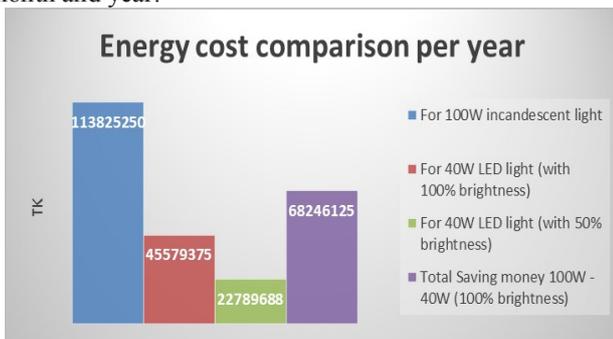

**Fig.9:** Cost comparison of three different streetlight systems in CCC.

According to our calculations, if LED light with 100% brightness burn for 12 hours, it can save approximately 60% electricity in compare to the conventional existing system in CCC [40.5MW - 16.2MW = 24.3MW, (24.3/40.5) x 100 = 60%] and hence saving 60% money as well. Since our system has a brightness control system, it is possible to save even 5% - 8% more on both electricity and money standards. Fig.9 illustrates the graphical representation of our cost comparison in CCC with a yearly saving of roughly 60 million TK just by replacing 100W sodium light to 40W LED in 100% brightness condition using the proposed system in this study. Study [19] reveals wastage of money worth 3.4 million TK per month also supports the results we get from our calculation.

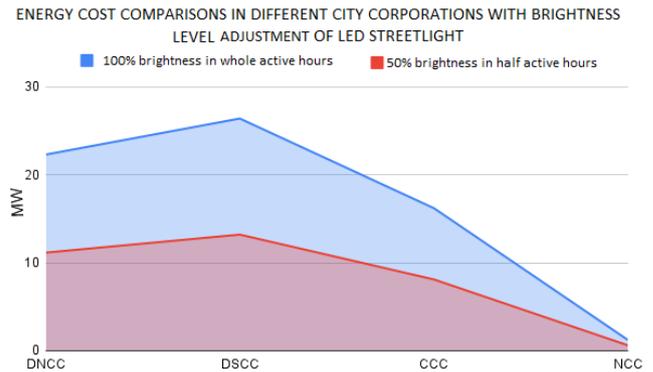

**Fig.10:** Cost comparison of four different city corporations in Bangladesh with different brightness combination of streetlights.

Combining the implementation of dimming condition during late night hours with full brightness condition during peak hours, the overall cost of energy in four city corporations in Bangladesh has significantly reduced (Fig.10) compare to the full brightness condition for the whole active hours that would be feasible for the country in tackling huge energy crisis in the post COVID-19 era of 2022.

We used an Arduino Nano in our project; however, we could develop it using a Raspberry Pi. We, on the other hand, used Arduino Nano to save money as it was developed to assess its compatibility. As well as in our experiment, we used an ultrasonic sensor, but an infrared sensor can deliver good results if the road distance is larger than the street light. Many people have utilized ZIGBEE to solve the same problem. The ZIGBEE, on the other hand, is useless for major cities if its communication range is less than 50 meters. Therefore, it has a favorable effect in small residential areas.

## VI. CONCLUSION

We presented and discussed the feasibility of IoT-based streetlight monitoring, controlling and error detecting system specially designed for major cities in Bangladesh. Our proposed system has demonstrated strong potential in tackling enormous electricity waste due to the existing improper and faulty streetlight system and provides a sophisticated way to monitor the whole city's streetlight management. With the adaptation of such system we can ensure to build a sustainable city in near future.

## REFERENCES

[1] "History of Street Lighting." *History of Street Lighting - Development of Street Lighting Technology*, http://www.historyoflighting.net/electric-lighting-history/history-of-street-lighting.

[2] "How Paris Became the City of Light: Holly Tucker." *Lapham's Quarterly*, https://www.laphamsquarterly.org/roundtable/how-paris-became-city-light.

[3] Lee, In, and Kyoochun Lee. "The Internet of Things (IOT): Applications, Investments, and Challenges for Enterprises." *Business Horizons*, Elsevier, 28 Apr. 2015, https://www.sciencedirect.com/science/article/pii/S0007681315000373.

[4] commission's report, B., 2022. [online] Berc.org.bd. Available at: <http://www.berc.org.bd/sites/default/files/files/berc.portal.gov.bd/notices/6d16474b_95b0_4b8a_aa9e_bad7767077eb/2020-02-27-16-12-6cbd095b3b66a47f99572580b080d036.pdf> [Accessed 27 February 2020].

[5] Hannan, S., Milton, G.B., Kabir, M.H. and Uddin, M.J., 2019, May. A Case Study on a Proposed Adaptive and Energy Efficient Street Lighting System for Chittagong City. In *2019 1st International Conference on Advances in Science, Engineering and Robotics Technology (ICASERT)* (pp. 1-5). IEEE.

[6] Hannan, S., Kabir, M.H., Milton, G.B. and Uddin, M.J., 2018, December. Design and Analysis of an Automatic and Adaptive Energy Efficient Street Lighting System for Chittagong Area. In *International Conference on Innovation in Engineering and Technology (ICIET)* (Vol. 27, p. 29).

[7] Alex, R. S., & Starbell, R. N. (2014). Energy efficient intelligent street lighting system using ZIGBEE and sensors. *International Journal of Engineering and Advanced Technology (IJEAT)*, *3*(4), 2249-8958.

[8] Bhairi, M. N., Kangle, S. S., Edake, M. S., Madgundi, B. S., & Bhosale, V. B. (2017, May). Design and implementation of smart solar LED street light. In *2017 International Conference on Trends in Electronics and Informatics (ICEI)* (pp. 509-512). IEEE.

[9] Attia, H.A., Omar, A. and Takruri, M., 2016, December. Design of decentralized street LED light dimming system. In *2016 5th International Conference on Electronic Devices, Systems and Applications (ICEDSA)* (pp. 1-4). IEEE.

[10] Khade, D.R., Gajane, N.V., Gawade, S.N. and Metri, R.A., 2017, April. Intensity controller of LED street lights. In *2017 International Conference on Circuit, Power and Computing Technologies (ICCPCT)* (pp. 1-4). IEEE.

[11] Menniti, D., Burgio, A. and Fedele, G., 2010, May. A cost-effective ac voltage regulator to mitigate voltage sags and dim lamps in street-lighting applications. In *2010 9th International Conference on Environment and Electrical Engineering* (pp. 396-399). IEEE.

[12] Velasco-Quesada, G., Roman-Lumbreras, M. and Conesa-Roca, A., 2011. Comparison of central dimmer systems based on multiple-tapped autotransformer and high-frequency switching converter. *IEEE Transactions on Industrial Electronics*, *59*(4), pp.1841-1848.

[13] Saifuzzaman, M., Moon, N. N., & Nur, F. N. (2017, December). IoT based street lighting and traffic management system. In *2017 IEEE region 10 humanitarian technology conference (R10-HTC)* (pp. 121-124). IEEE.

[14] Sakib Ullah Sourav, M. and Wang, H., 2022. Energy Efficient Automatic Streetlight Controlling System using Semantic Segmentation. *arXiv e-prints*, pp.arXiv-2209.

[15] Department of Engineering-Electrical, "Chattogram City Corporation," 2018. http://www.ccc.org.bd/electrical.[Accessed: 2018- 09- 20].

[16] The Financial Express. "DNCC to Complete Installing 46,410 LED Lights by 2021." *The Financial Express*, https://today.thefinancialexpress.com.bd/metro-news/dncc-to-complete-installing-46410-led-lights-by-2021-1599582731.

[17] "Dhaka South City Corporation." http://dscc.gov.bd/site/page/c918f990-1f75-46cb-95ad-c08d10197af5/%E0%A6%8F%E0%A6%95-%E0%A6%A8%E0%A6%9C%E0%A6%B0%E0%A7%87-%E0%A6%A2%E0%A6%BE%E0%A6%95%E0%A6%BE-%E0%A6%A6%E0%A6%95%E0%A7%8D%E0%A6%B7%E0%A6%BF%E0%A6%A3.

[18] "Narayanganj City Corporation." https://ncc.gov.bd/site/page/d7dd6dd2-d697-4620-976e-ede6b3485e7f/%E0%A6%8F%E0%A6%95-%E0%A6%A8%E0%A6%9C%E0%A6%B0%E0%A7%87-%E0%A6%B8%E0%A6%BF%E0%A6%9F%E0%A6%BF-%E0%A6%95%E0%A6%B0%E0%A7%8D%E0%A6%AA%E0%A7%8B%E0%A6%B0%E0%A7%87%E0%A6%B6%E0%A6%A8?fbclid.

[19] "Much Power Gets Wasted in Streetlamps." *Dhaka Tribune*, 23 Aug. 2014,https://archive.dhakatribune.com/uncategorized/2014/08/23/much-power-gets-wasted-in-streetlamps.

## Appendix I

Scan this QR code to watch the report we made based on highly credible news articles, media reports on television and our on spot survey.

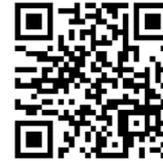